\documentclass[aps,prl,twocolumn,showpacs,superscriptaddress]{revtex4}
\usepackage{graphicx,amssymb,amsmath,epsf,bm}
\epsfclipon

\newcommand{\ri}{\mathrm{i}}
\newcommand{\re}{\mathrm{e}}
\newcommand{\ep}{\varepsilon}
\renewcommand{\(}{\left(}
\renewcommand{\)}{\right)}

\newcommand{\rd}{\mathrm{d}}

\newcommand{\expct}[1]{\langle #1 \rangle}

\newcommand{\const}{\text{cst.}}  
\newcommand{\prt}[2]{\frac{\partial #1}{\partial #2}}
\newcommand{\prts}[3]{\frac{\partial^{#3} #1}{\partial {#2}^{#3}}}

\renewcommand{\eqref}[1]{Eq.~(\ref{#1})}

\newcommand{\pref}[1]{(\ref{#1})}
\newcommand{\figref}[1]{Fig.~\ref{#1}}

\begin{document}

\title{Extensive and Sub-Extensive Chaos in Globally-Coupled Dynamical Systems}

\author{Kazumasa A. Takeuchi}
\affiliation{Service de Physique de l'\'Etat Condens\'e,~CEA -- Saclay,~F-91191~Gif-sur-Yvette,~France}%

\author{Hugues Chat\'e}
\affiliation{Service de Physique de l'\'Etat Condens\'e,~CEA -- Saclay,~F-91191~Gif-sur-Yvette,~France}%

\author{Francesco Ginelli}
\affiliation{Dipartimento di Fisica and INFN, Universit\'a di Firenze,
Via Sansone 1, I-50019 Sesto Fiorentino, Italy}
\affiliation{Service de Physique de l'\'Etat Condens\'e,~CEA -- Saclay,~F-91191~Gif-sur-Yvette,~France}%

\author{Antonio Politi}
\affiliation{CNR, Istituto dei Sistemi Complessi, via Madonna del Piano 10, I-50019 Sesto Fiorentino, Italy}

\author{Alessandro Torcini}
\affiliation{CNR, Istituto dei Sistemi Complessi, via Madonna del Piano 10, I-50019 Sesto Fiorentino, Italy}
\affiliation{Dipartimento di Fisica and INFN, Universit\'a di Firenze,
Via Sansone 1, I-50019 Sesto Fiorentino, Italy}

\date{\today}




\begin{abstract}
Using a combination of analytical and numerical techniques, we show that chaos
in globally-coupled identical dynamical systems,
 be they dissipative or Hamiltonian, is 
both extensive and sub-extensive: 
their spectrum of Lyapunov exponents 
is asymptotically flat (thus extensive) at the value $\lambda_0$ given by a 
single unit forced by the mean-field,
but sandwiched between sub-extensive bands containing typically 
$\mathcal{O}(\log N)$ exponents whose values vary as
 $\lambda \simeq \lambda_\infty + c/\log N$
 with $\lambda_\infty \neq \lambda_0$.
\end{abstract}


\maketitle

Dynamical systems made of many coupled units
 with long-range or global coupling are models
 of numerous important situations in physics and beyond,
  ranging from the synchronization of oscillators and neural networks
 to gravitational systems, plasma, and hydrodynamics
\cite{Dauxois_etal-Book2002,Kura-book}.
Their properties can be quite remarkable:
 for instance,
 globally-coupled dissipative systems can
 give rise to collective chaos,
 where macroscopic variables show incessant irregular behavior
 due to nontrivial correlations between local units
 \cite{Collective_chaos}.
Their Hamiltonian counterparts, in the microcanonical ensemble,
 are now well-known to show negative specific heat,
 long-lived quasi-stationary states,
 all features ultimately related to their non-additivity
 \cite{Dauxois_etal-Book2002}. 
Their unusual properties make these systems deceptively close to 
simple mean-field approximations and show that they are, in many ways, 
less well understood than systems with short-range interactions.

The status of the chaos
which may be present in these dynamical systems
is particularly unclear:
whereas systems with short-range interactions are now well-known to exhibit 
 extensive chaos \cite{Extensivity}, at least in the absence 
of non-trivial collective behavior \cite{Takeuchi_etal-PRL2009},
 there is, in our view,
 no solid evidence or argument for or against the extensivity of chaos
 in long-range or globally-coupled systems.

\begin{figure}[t]
 \includegraphics[width=\hsize,clip]{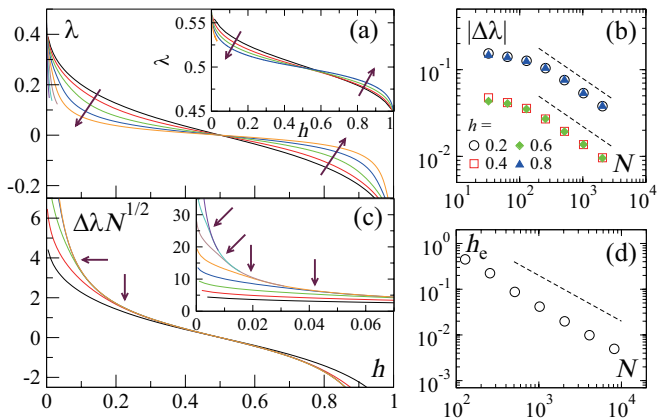}
 \caption{(color online). Full Lyapunov spectrum \cite{NUM}.
(a) Spectra for different sizes (arrows: increasing $N$) 
for the RM model with $a=1$ 
(main panel,  $N=128, 256, 512, \dots$) and for the skewed-tent GCM with $b=4$ 
(inset, $N=32, 128, 512, \dots$). 
(b) $|\Delta\lambda|$ vs $N$ at fixed values of $h$. 
Dashed lines: $|\Delta\lambda| \sim 1/\sqrt{N}$. 
(c) Same data as in main panel of (a) in rescaled coordinate from $N = 128$
(main panel, innermost curve) to $N=16384$ (inset, outermost curve). 
Arrows indicate the positions $h_{\rm e}(N)$ at which spectra of size $N$ and $2N$ start to collapse. (d) $h_{\rm e}(N)$ vs $N$. Dashed line: $h_{\rm e}(N) \sim 1/N$.}
 \label{fig:FullSpec}
\end{figure}

In most dynamical systems, chaos is customarily quantified by 
Lyapunov exponents (LEs), which measure the average rate of divergence 
of nearby trajectories, and more specifically by Lyapunov spectra, 
where the LEs $\lambda^{(i)}$ are arranged in descending order.
When Lyapunov spectra, plotted as functions of $(i- \frac{1}{2})/N$, 
collapse reasonably well onto each other for different system sizes $N$,
chaos is deemed extensive.
This has been observed repeatedly in 
locally-coupled systems, but never with global or long-range coupling,
even for the largest sizes reachable numerically today 
 (see, e.g., 
\figref{fig:FullSpec}(a) below).
Yet, a na\"{\i}ve argument suggests extensivity in this latter case:
for identical units submitted
 to the same self-consistent mean-field forcing,
the LEs should all take the same value 
(hereafter $\lambda_0$),
 a trivial realization of extensivity. 
But this is strictly true only if the $N\to\infty$ limit is taken first, 
which may be misleading when dealing with LEs, as they are essentially 
infinite-time averages. In fact,
the Lyapunov spectra of finite-size globally-coupled systems always remain 
far from being flat.

For the paradigmatic and much-studied Hamiltonian mean-field (HMF) model
 \cite{Antoni_Ruffo-PRE1995,Dauxois_etal-LNP2002},
 the situation is similarly confusing:
 the na\"{\i}ve argument above
gives all LEs at zero,
 whereas a calculation by Firpo yielded a positive largest exponent 
at any finite $N$, with a well-defined $N\to\infty$ limit \cite{Firpo-PRE1998}.
A theoretical
 formulation as a quantum many-body problem 
mentioned the possibility of a vanishing fraction of non-zero exponents
\cite{TanaseKurchan-JPhysA2003},
 but numerical results have produced contradictory results
 \cite{Dauxois_etal-LNP2002,Others}.

In this Letter, we show that chaos is {\it not} fully extensive
in systems of globally-coupled identical units.
Rather, their Lyapunov spectra, in the large-size limit,
 converge to flat extensive regions where the LEs do take the 
value $\lambda_0$ given by the single unit forced by the mean field,
but these regions are bordered by sub-extensive layers
containing exponents taking different values.
In particular, we provide a theoretical analysis and numerical evidence 
showing that the largest LE $\lambda^{(1)}$ converges
 as $\lambda^{(1)} \simeq \lambda_\infty + c / \log N$
to an asymptotic value $\lambda_\infty >\lambda_0$.
Our numerical analysis reveals that the sub-extensive boundary layers 
 contain $\mathcal{O}(\log N)$ Lyapunov modes
 and that their LEs take the same asymptotic value $\lambda_\infty$.
We finally argue that our results probably hold also in the presence of collective chaos.

We first study $N$ globally-coupled dissipative maps
\begin{equation}
 x_j^{t+1} = f(y_j^t),~~~~ y_j^t = (1-\ep) x_j^t + \frac{\ep}{N} \sum_{j'=1}^N x_{j'}^t,  \label{eq:GCM}
\end{equation}
 with $j = 1, \dots, N$, time $t$, coupling constant $\ep$,
 and a chaotic local map $f(x)$, which is chosen here to be one-dimensional
 for the sake of simplicity.
If $f(x)$ shows sufficiently strong mixing,
 its Jacobian may be approximated by a random multiplier.
The tangent-space dynamics of \eqref{eq:GCM} is then simplified as
\begin{equation}
 v_j^{t+1} = \mu_j^t \biggl[ (1-\ep) v_j^t + \frac{\ep}{N} \sum_{j'=1}^N v_{j'}^t \biggr],  \label{eq:RM}
\end{equation}
 with iid random numbers $\mu_j^t$, unless the coupling $\ep$ is too strong
 to regard $f'(y_j^t)$ as independent.
The mean-field forcing argument amounts to ignoring
 the global-coupling term in \eqref{eq:RM}, which is then
reduced to the biased Brownian motion of a particle of coordinate $\log |v_j^t|$
 with average velocity $\lambda_0 \equiv \expct{\log|(1-\ep)\mu_j^t|}$
 and diffusion coefficient
 $D \equiv \expct{(\log|(1-\ep)\mu_j^t| - \lambda_0)^2}$,
where $\lambda_0$ is 
the mean-field LE.
From this viewpoint, the full system \pref{eq:RM} can be seen as
 $N$ interacting Brownian particles.
Assume now that the Lyapunov vector $[v_1^t, \dots, v_N^t]$
 is sufficiently localized,
 which is indeed the case except when it is associated with
 collective behavior \cite{Takeuchi_etal-PRL2009}.
In this case, its largest component $v_M^t$ dominates
 the coupling term in \eqref{eq:RM}.
Thus, the Brownian particles $\log |v_j^t|$ diffuse freely
as long as $|(1-\ep) v_j^t| \gg |(\ep/N)v_M^t|$, otherwise
the coupling term takes effect, keeping any $|(1-\ep)v_j^t|$ larger
 than $|(\ep/N)v_M^t|$.
In other words, the $N$ Brownian particles $\log |v_j^t|$
 diffuse within a box of size $\log [N(1-\ep)/\ep]$,
 whose right end corresponds to the rightmost particle,
 while the other end pulls all the particles left behind.
The first LE $\lambda^{(1)}$ is then simply given
 as the average velocity of this box.
This process is described
 by the following Fokker-Planck equation
 in a frame moving at velocity $\lambda^{(1)}$:
\begin{equation}
 \prt{}{t}P(u,t) = -\prt{}{u}[(\lambda_0 - \lambda^{(1)}) P] + \frac{D}{2}\prts{P}{u}{2},  \label{eq:FP}
\end{equation}
 where $u$ is the coordinate in this frame
 and the particle distribution function $P(u,t)$
 is confined, roughly,
 in $0 \leq u \leq u_{\rm max} \equiv \log [N(1-\ep)/\ep]$.
For large $N$, its stationary solution can be approximated
 by the one in the limit $u_{\rm max}\to\infty$,
 $P_{\rm s}(u) = (2\Delta\lambda^{(1)}/D) \exp(-2\Delta\lambda^{(1)} u/D)$
 with $\Delta\lambda^{(1)} \equiv \lambda^{(1)} - \lambda_0$.
Further, by the definition of the box,
 there should be
 $\mathcal{O}(1)$ particles near its right end $u_{\rm max}$,
 which implies
 $\int_{u_{\rm max}}^\infty P_{\rm s}(u) \rd u = c_1/N$
 with a constant $c_1 \sim \mathcal{O}(1)$.
This yields our central result for the first LE:
\begin{equation}
 \Delta\lambda^{(1)}
 = \lambda^{(1)} \!-\! \lambda_0
 = \frac{D}{2}\!\(1 \!+\! \frac{c_2}{\log N} \) \!+\! \mathcal{O}\( \frac{1}{\log^2 N} \),  \label{eq:MaxLE}
\end{equation}
 with
$c_2 \equiv \log [\ep/((1-\ep)c_1)]$.
The probability distribution $\mathcal{P}(v)$
 for the vector components $v_j$
 is $\mathcal{P}(v) = P_{\rm s}(\log v) (\rd u/\rd v) \sim v^{-2-c_2/\log N}$,
 whose exponent is smaller than $-1$ and thus consistent with
 our assumption of localization of the Lyapunov vector.
A similar result holds for the {\it last} LE \cite{TBP}:
$\Delta\lambda^{(N)} \equiv \lambda^{(N)} - \lambda_0 \simeq -(D/2)(1 + c'_2/\log N)$ with another coefficient $c'_2$.

\begin{figure}[t]
 \includegraphics[width=\hsize,clip]{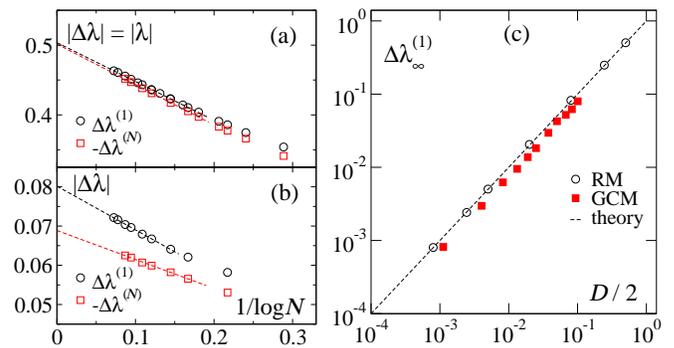}
 \caption{(color online). Size-dependence of the first and last LEs \cite{NUM}.
(a,b): $\Delta\lambda^{(1)}$ and $|\Delta\lambda^{(N)}|$ against $1/\log N$ 
for the RM model with $a=1$ (a) and for the skewed-tent GCM with $b=4$ (b). 
Dashed lines indicate linear fits to the data. 
(c) Estimated value of $\Delta\lambda_\infty^{(1)} = \lim_{N\to\infty}\Delta\lambda^{(1)}$ 
for the RM model and our GCM with varying $a$ and $b$, respectively, 
plotted against $D/2$. The diffusion constant $D$ is obtained by $D=a^2$ 
for the RM model and numerically measured for the GCM. 
Dashed line: $\Delta\lambda_\infty^{(1)}=D/2$ as predicted in \eqref{eq:MaxLE}. 
}
 \label{fig:MaxLE}
\end{figure}%

These results are confirmed in \figref{fig:MaxLE}
 by direct simulations of the random multiplier (RM) model \pref{eq:RM}
 and of globally-coupled maps (GCM) \pref{eq:GCM} \cite{NUM}.

For the RM model,
 we used $\ep = 0.1$ and $\mu_j^t = \pm \exp{\xi_j^t}/(1-\ep)$
 with random signs (here ``$+$'' with probability $0.6$) 
and $\xi_j^t$ drawn from the centered Gaussian with variance $a^2$, 
which gives $\lambda_0 = 0$ and $D = a^2$.
{\it Quantitative} agreement is found with
\eqref{eq:MaxLE} for the first LE and
 its counterpart for the last LE [\figref{fig:MaxLE}(a,c)].

For our GCM system, we chose skewed tent maps
$f(x) = bx$ (resp. $ b(x-1)/(1-b)$) if $0 \leq x \leq 1/b$ (resp. $1/b < x \leq 1$)
 coupled with strength $\ep = 0.02$.
The results in \figref{fig:MaxLE}(b)
 demonstrate again the logarithmic size-dependence
 of $\Delta\lambda^{(1)}$ and $\Delta\lambda^{(N)}$.
Their asymptotic values are not symmetric anymore [\figref{fig:MaxLE}(b)],
 but the deviation from $D/2$ remains small [\figref{fig:MaxLE}(c)].
In addition, we note that here 
 $\lambda_0$
depends residually on $N$ through changes in the invariant measure.
This effect is however so weak
that in practice we observe the same logarithmic law
 for the first and last LEs, $\lambda^{(1)}$ and $\lambda^{(N)}$.

Let us summarize our results so far:
The first and last LEs remain distinct
from the mean-field forcing LE $\lambda_0$ in the $N\to\infty$ limit.
They are shifted from $\lambda_0$ by an amount $\Delta\lambda$ controlled by $D$,
 the amplitude of the fluctuations in the Jacobian,
and the coupling strength $\ep$ is only involved in the logarithmic 
finite-size corrections \cite{Daido}.

We now investigate the full Lyapunov spectrum of our systems.
As seen above, it cannot be entirely flat
at $\lambda_0$ asymptotically.
However, for finite-size systems, Lyapunov spectra
 become flatter for larger sizes under the conventional rescaling
 $\lambda^{(i)}$ vs $h \equiv (i- \frac{1}{2})/N$
 [\figref{fig:FullSpec}(a)].
In the RM model, a closer look at the ``bulk'' LEs with fixed $h$
 reveals an asymptotic power-law decay
 $\Delta\lambda(h) \equiv \lambda^{(i)} - \lambda_0 \sim 1/\sqrt{N}$
 toward the mean-field forcing value
 $\lambda_0 = 0$ [\figref{fig:FullSpec}(b)]. 
This scaling is only reached for large-enough sizes and sooner near the
middle of the spectrum, as shown clearly by
rescaled spectra $\Delta\lambda\,\sqrt{N}$ [\figref{fig:FullSpec}(c)]:
they collapse very well within a central region $[h_{\rm e}(N), 1-h_{\rm e}(N)]$,
 with $h_{\rm e}(N)$ decreasing toward zero as $1/N$
 [\figref{fig:FullSpec}(c) arrows and \figref{fig:FullSpec}(d)].
Thus, in the infinite-size limit,
the Lyapunov spectrum of the RM model is indeed flat
at  $\lambda_0$,
but sandwiched between two sub-extensive bands of LE taking 
different values  \cite{NOTE}.

\begin{figure}[t]
 \includegraphics[width=\hsize,clip]{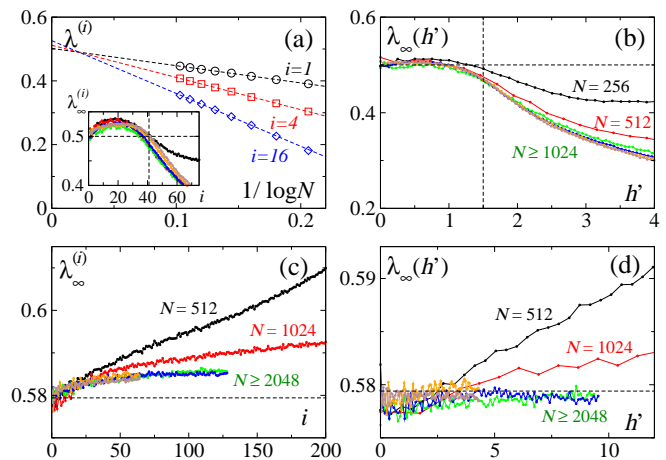}
 \caption{(color online). Left sub-extensive band of the Lyapunov spectrum for the 
RM model with $a=1$ (a,b) and for the skewed-tent GCM with $b=4$ (c,d) \cite{NUM}. 
(a) $\lambda^{(i)}$ vs $1/\log N$ for $i=1,4,16$. 
Dashed lines indicate linear fits to the data. 
Inset: $\lambda_\infty^{(i)} \equiv (\lambda_{2N}^{(i)}\log 2N - \lambda_N^{(i)}\log N) / \log 2$
vs $i$ for $N = 256, 512, \dots, 8192$ (see text). 
The horizontal and vertical dashed lines indicate $\lambda_\infty^{(1)}$ and a
threshold index value estimated from $h_{\rm e}$, respectively.
(b) Same plot as the inset of (a) but with rescaled indices 
$h' \equiv (i-1)/(i_0+\log N)$ with $i_0=15$. 
(c,d) Same plots as the inset of (a) and (b) for the GCM with 
$N = 512, 1024, \dots, 16384$, and $i_0=5$.}
 \label{fig:SubExt}
\end{figure}%

That $h_{\rm e} \sim 1/N$ [\figref{fig:FullSpec}(d)]
implies that the number of non-extensive LEs
 increases slower than any power of $N$.
We now show that it actually grows logarithmically with $N$.
With fixed indices $i$,
 these LEs at size $N$ seem to obey \eqref{eq:MaxLE},
 $\lambda_N^{(i)} \simeq \lambda_\infty^{(i)} + c^{(i)}/\log N$
 [\figref{fig:SubExt}(a)],
 but the estimated
 $\lambda_\infty^{(i)}$ increase with $i$ (dashed lines),
 at odds with the monotonicity of the Lyapunov spectrum.
This is better seen when plotting
 $(\lambda_{2N}^{(i)}\log 2N - \lambda_N^{(i)}\log N) / \log 2$
 as estimates for $\lambda_\infty^{(i)}$
 [inset of \figref{fig:SubExt}(a)],
 where $\lambda_\infty^{(i)}$ is found to be larger
 than $\lambda_\infty^{(1)}$ within the non-extensive region
 $1 \leq i \lesssim i_{\rm e} \equiv h_{\rm e}N$.
Instead, if we rescale the index logarithmically
 as $h' \equiv (i-1)/(i_0+\log N)$,
 with $i_0$ adjusted here for the LEs to show the $1/\log N$ law,
 the asymptotic LEs $\lambda_\infty(h')$ do not increase
 with $h'$ anymore, but stay constant
 in the non-extensive region except
 near the threshold (\figref{fig:SubExt}(b), left of the dashed line).
This indicates that all the non-extensive LEs
 converge to the same value as the first LE
 and that their number increases logarithmically with $N$.
The same conclusion is reached for our GCM system
 [\figref{fig:SubExt}(c,d)],
 though we could not compute all the non-extensive LEs
 within reasonable time \cite{NUM}.
In short, we find that $\mathcal{O}(N)$ extensive LEs
 are sandwiched by two sub-extensive bands
 at both ends of the spectrum,
 each of which consists of $\mathcal{O}(\log N)$ LEs
 with asymptotic values shifted approximately by $D/2$
 from $\lambda_0$.

We now show that our results also extend to the HMF model, and 
thus probably also to other globally-coupled Hamiltonian models.
Defined by the Hamiltonian
 $H = \frac{1}{2}\sum_j p_j^2 + \frac{1}{2N} \sum_{j,j'} [1-\cos(\theta_j - \theta_{j'})]$, the HMF model
is intensely studied mostly because its infinite-size limit displays
an abundance of non-trivial solutions which appear as so-called 
quasi-stationary states at finite $N$  
\cite{Antoni_Ruffo-PRE1995,Dauxois_etal-LNP2002}.
Contradictory results exist about the nature
of chaos in this model
 \cite{Dauxois_etal-LNP2002,Others},
 even in its reference ``equilibrium'' state.
The motion of a single particle is given by
 $\ddot{\theta}_j = -M\sin(\theta_j - \Theta)$,
 where $M\re^{\ri\Theta} \equiv \frac{1}{N}\sum_j \re^{\ri\theta_j}$ is the 
mean field which is non-zero in the (equilibrium)
ferromagnetic phase present for 
energy density $U < \frac{3}{4}$.
Here the na\"ive argument
yields $\lambda_0=0$ because a single particle
forced by a constant mean-field cannot be chaotic. 

\begin{figure}[t]
 \includegraphics[width=\hsize,clip]{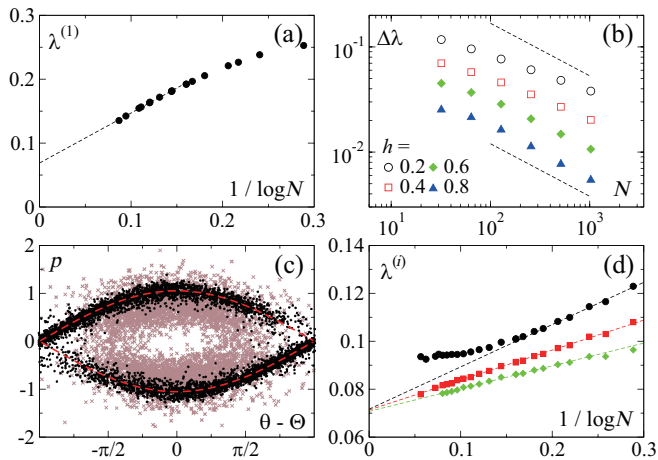}
 \caption{(color online) (a-c): HMF model with energy density $U=0.7$ \cite{NUM}. 
(a) First LE, maximum size $N=10^5$. Dashed line: linear fit in the $1/\log N$ regime. 
(b) $\Delta\lambda$ ($=\lambda$) vs $N$ at fixed rescaled indices $h$. Dashed lines: $\Delta\lambda \sim 1/\sqrt{N}$.
(c) Positions of the oscillator most contributing to the 1st and 256th Lyapunov vector (black dots and brown crosses, respectively) at $N=512$. Red dashed line indicates the separatrix.
(d) First 3 LEs for globally-coupled Hopf oscillators (see text).
}
 \label{fig:HMF}
\end{figure}%

We were able to extend the argument leading to \eqref{eq:MaxLE} 
to the HMF model (details will appear elsewhere
\cite{TBP}): At finite $N$, the mean 
field fluctuates and the energy of a particle diffuses, so that 
it eventually visits the surroundings of  $U_M=1+M$,
the unstable maximum of the mean-field potential. There, it
experiences  a chaotic kick, and this produces 
a finite diffusion coefficient $D$ for the 
logarithm of the tangent-space amplitudes.
Taking these effects into account, we obtain a strictly positive 
asymptotic first LE with, again, $1/\log N$ corrections, which we
numerically confirm in \figref{fig:HMF}(a). 
The full Lyapunov spectrum, on the other hand,
gets flatter for larger $N$ and we could observe
the emergence of the $\Delta\lambda(h)\sim 1/\sqrt{N}$ 
scaling [\figref{fig:HMF}(b)], but
we are currently unable
to study the larger systems,
 in order to overcome finite-size effects to
 obtain clear evidence of ${\mathcal O}(\log N)$ sub-extensive LEs.
Nevertheless, it is already clear from \figref{fig:HMF}(b) 
that the $h$-domain where
the $1/\sqrt{N}$ scaling holds widens with $N$, suggesting
a flat (zero-valued) extensive part with a sub-extensive, 
possibly logarithmic, band of positive LEs.

We finally examine the influence of collective chaos
on our results, an important generic case for dissipative globally-coupled systems
\cite{Collective_chaos}.
Lyapunov spectra then contain modes governing the macroscopic
dynamics, whose associated covariant Lyapunov vectors are delocalized 
\cite{Takeuchi_etal-PRL2009}.
In the case of globally-coupled limit-cycle oscillators,
$\dot{W}_j = W_j - (1+\mathrm{i}c_2)|W_j|^2 W_j + K (1+\mathrm{i}c_1) (\expct{W} - W_j)$,
 with complex variables $W_j$, $\expct{W} \equiv \frac{1}{N}\sum_j W_j$,
 $c_1 = -2.0$, $c_2 = 3.0$, and $K = 0.47$,
the largest LE is such a collective mode \cite{Takeuchi_etal-PRL2009}. Here we see
that it does not obey \eqref{eq:MaxLE} for $N$ large enough
while the \textit{following}, ``non-collective'' LEs do  [\figref{fig:HMF}(d)].
This result comforts the general picture of the macroscopic modes being present
but asymptotically decoupled from the other ones in Lyapunov spectra, be they in the bulk or
in the subextensive layers.

Our findings recall the importance of the order of limits in systems with long-range interactions:
For the ferromagnetic phase of the HMF model for instance,
considering directly the infinite-size system
 (the Vlasov equation \cite{Dauxois_etal-LNP2002}), one misses the fact that 
residual but influential chaos remains in the $N\to\infty$ limit,
even though the bulk exponents vanish asymptotically.
The Lyapunov modes in the subextensive layers capture ``extreme events'' in  phase space, much like 
the largest LE in locally-coupled systems \cite{Lyapsti,Egolf}. 
For the HMF case, this is particularly clear since,
 whereas the covariant vectors for bulk LEs are carried by typical oscillators,
 the first Lyapunov vector is localized
 on those oscillators currently in the vicinity of the separatrix,
 the most unstable part of (local) phase space
 [\figref{fig:HMF}(c)].
Even though they are in logarithmic numbers and localized on special regions
of phase space, the Lyapunov modes of the subextensive layers may have an important impact on 
macroscopic properties, such as the thermodynamic entropy of Hamiltonian systems \cite{entropy}.

In summary, we have shown
 that microscopic chaos in systems made of $N$ globally-coupled dynamical units
 exhibits a rather peculiar form of extensivity:
 their Lyapunov spectrum $\lambda(h)$ is asymptotically flat, 
thus ``trivially'' extensive,
 but sandwiched between sub-extensive bands with LEs
 taking different values.
In presence of macroscopic dynamics, the corresponding collective Lyapunov modes
are just superimposed on this structure.
The bulk LEs converge as $\lambda(h) \simeq \lambda_0 +\const/\sqrt{N}$
to the value $\lambda_0$
 given by a single dynamical unit forced by the mean-field.
In contrast, the sub-extensive layers
 contain $\mathcal{O}(\log N)$ LEs
 whose values vary as
 $\lambda \simeq \lambda_\infty + \const/\log N$
 with $\lambda_\infty \neq \lambda_0$.
Investigating further the genericity of our results and 
providing a theoretical basis
to the $1/\sqrt{N}$ scaling of bulk LEs and the $\log N$ size of 
sub-extensive bands are important tasks left for future study.


\begin{thebibliography}{99}
\bibitem{NOTE}
For GCM, the scaling $\Delta\lambda \sim 1/\sqrt{N}$
 for the bulk LEs is not fully established
at the largest size we could reach.
\end{thebibliography}

\end{document}